\documentclass[amssymb,prd,superscriptaddress,aps,nofootinbib,twocolumn,showpacs]{revtex4-1}
\usepackage{graphicx, epsfig, amssymb} 
\usepackage{amsmath, amsfonts}
\usepackage{bm} 
\usepackage[breaklinks]{hyperref}
\usepackage{color}
\usepackage{enumerate}

\def\nn{\nonumber}

\def\be{\begin{equation}}
\def\ee{\end{equation}}
\def\beq{\begin{eqnarray}}
\def\eeq{\end{eqnarray}}

\begin{document}

\title{I-Love-Q relations for a gravastar and the approach to the black-hole limit}

\author{Paolo Pani}\email{paolo.pani@roma1.infn.it}
\affiliation{Dipartimento di Fisica, ``Sapienza'' Universit\`a di Roma \& Sezione INFN Roma1, Piazzale Aldo Moro 5, 00185, Roma, Italy.}
\affiliation{CENTRA, Departamento de F\'{\i}sica, Instituto Superior T\'ecnico, Universidade de Lisboa, Avenida~Rovisco Pais 1, 1049 Lisboa, Portugal.}

\begin{abstract} 
The multipole moments and the tidal Love numbers of neutron stars and quark stars satisfy certain relations which are almost insensitive to the star's internal structure. A natural question is whether the same relations hold for different compact objects and how they possibly approach the black-hole limit. Here we consider ``gravastars'', which are hypothetical compact objects sustained by their internal vacuum energy.
Such solutions have been proposed as exotic alternatives to the black-hole paradigm because they can be as compact as black holes and exist in any mass range.
By constructing slowly-rotating, thin-shell gravastars to quadratic order in the spin, we compute the moment of inertia $I$, the mass quadrupole moment $Q$, and the tidal Love number $\lambda$ in exact form. 
The $I$-$\lambda$-$Q$ relations of a gravastar are dramatically different from those of an ordinary compact star, but the black-hole limit is continuous, i.e. these quantities approach their Kerr counterparts when the compactness is maximum. 
Therefore, such relations can be used to discern a gravastar from an ordinary compact star, but not to break the degeneracy with the black-hole case.
Based on these results, we conjecture that the full multipolar structure and the tidal deformability of a spinning, ultracompact gravastar are identical to those of a Kerr black hole.
The approach to the black-hole limit is nonpolynomial, thus differing from the critical behavior recently found for strongly anisotropic neutron stars.
\end{abstract}

\pacs{
04.20.-q,  	
04.25.-g,	
04.70.Bw,	
04.30.-w.	
}

\maketitle

\section{Introduction}
Astrophysical black holes (BHs) are the simplest macroscopic objects in the Universe, being characterized only by their mass $M$ and angular momentum $J$. This property is formally proved by the no-hair and uniqueness theorems~\cite{Carter71,Hawking:1973uf,Heusler:1998ua,Chrusciel:2012jk,Robinson} which --~roughly speaking~-- state that any regular BH cannot possess further asymptotic charges other than $M$ and $J$, and that any stationary, vacuum solution of Einstein's equations is described by the Kerr metric. This ``two-hair'' property implies that all multipole moments of a Kerr BH can be written in terms of $M$ and $J$ in an elegant and compact form~\cite{Hansen:1974zz}.
The simplicity of BHs survives also in the presence of external tidal fields. The tidal Love numbers (which measure the tidal deformability of a self-gravitating object~\cite{Murraybook,PoissonWill}) of nonspinning~\cite{Binnington:2009bb,Damour:2009vw,Gurlebeck:2015xpa} and spinning~\cite{Poisson:2014gka,Pani:2015hfa,Landry:2015zfa} BHs are precisely zero\footnote{At least in the axisymmetric case to second order in the spin~\cite{Pani:2015hfa} and generically to first order in the spin~\cite{Landry:2015zfa}.}.

It has been recently discovered that relativistic compact stars satisfy certain nearly-universal relations which are reminiscent of the no-hair properties of BHs. The multipole moments and the tidal Love numbers of a neutron star (NS) depend on each other through relations which are almost insensitive to the star's internal structure, i.e. they depend only mildly (at the percent level) on the equation of state of matter in the NS interior~\cite{Yagi:2013bca,Yagi:2013awa} (cf. also Refs.~\cite{Lattimer:2000nx,Tsui:2004qd,Urbanec:2013fs} for some earlier related work).

In their simplest form these relations involve the moment of inertia $I$, the mass quadrupole moment $Q$, and the electric, quadrupolar tidal Love number $\lambda$. For this reason, they were dubbed ``I-Love-Q'' relations~\cite{Yagi:2013bca,Yagi:2013awa}.  
Although the I-Love-Q relations were originally discovered for
slowly-rotating, barotropic, isotropic, unmagnetized and isolated stars~\cite{Yagi:2013bca,Yagi:2013awa}, they have been extended to include rapid rotation~\cite{Doneva:2013rha,Pappas:2013naa,Chakrabarti:2013tca,Yagi:2014bxa}, nonbarotropic~\cite{Martinon:2014uua} and anisotropic~\cite{Yagi:2015cda,Yagi:2015hda} fluids, strong magnetic fields~\cite{Haskell:2013vha}, dynamical configurations~\cite{Maselli:2013mva} and also deviations from General Relativity~\cite{Yagi:2013bca,Yagi:2013awa,Sham:2013cya,Pani:2014jra,Doneva:2014faa}. The outcome of these studies is that the approximate universality is remarkably robust in realistic configurations.

It is therefore important to explain the origin of these properties. Some possible explanations were proposed in~\cite{Yagi:2014qua,Sham:2014kea} in terms of
an emergent approximate self-similarity in isodensity contours inside the star and in terms of the approach to the incompressible limit of realistic equations of state, respectively. Some analytical models to support these arguments were developed in Refs.~\cite{Stein:2014wpa,Chatziioannou:2014tha} and Ref.~\cite{Chan:2014}, respectively.

Two natural questions arising in this context are whether the same nearly-universal relations exist for compact objects other than NSs and quark stars, and whether such approximate universality is related to the approach to the BH limit for very compact objects. In this paper we wish to consider these two problems.

A major obstacle in understanding the approach to the BH limit consists in the fact that ordinary compact stars have a maximum mass and a maximum compactness and, therefore, cannot be as compact as a BH. In order to circumvent this problem, Refs.~\cite{Yagi:2015cda,Yagi:2015hda} recently studied the I-Love-Q relations for anisotropic, incompressible stars, which can almost reach the BH limit for large values of the anisotropic parameter. They found that the approach to the BH limit is continuous and reminiscent of a phase transition, with the multipole moments displaying a certain critical behavior.

In this paper, we study the I-Love-Q relations for a different model of compact object, namely for ``gravitational condensate stars'', or gravastars~\cite{Mazur:2001fv}.
In these models, the spacetime is assumed to undergo a quantum phase transition in the vicinity of the would-be event horizon. The horizon is effectively replaced by a transition layer and the BH interior is replaced by a
patch of de Sitter space~\cite{Mazur:2004fk}. The effective negative pressure of the de Sitter interior contributes to sustain the self-gravity of the object for any compactness. In the static case these models are thermodynamically~\cite{Mazur:2001fv} and dynamically~\cite{Visser:2003ge,Chirenti:2007mk,Pani:2009ss} stable to linear order\footnote{The arguments presented in Refs.~\cite{Keir:2014oka,Cardoso:2014sna} suggest that ultracompact objects without an event horizon might be \emph{nonlinearly} unstable. Furthermore, if highly spinning these objects are linearly unstable against the ergoregion instability~\cite{1978CMaPh..63..243F,Cardoso:2007az} (cf. Ref.~\cite{Brito:2015oca} for a review).} for reasonable equations of state of the transition layer.

In the context of the approximate universal relations mentioned above, gravastars have various interesting features. First, these solutions can exist for any compactness, smoothly connecting the Newtonian regime to the BH limit. As we will show, the model is sufficiently simple to admit an \emph{exact} solution, so that the I-Love-Q relations for a gravastar can be computed analytically. This also sheds some light on the approach to the BH limit, which is not polynomial in this model but involves logarithmic terms. Secondly, in general there is no reason to expect that the BH limit of a compact material body be continuous and one could hope to use the I-Love-Q relations to break the degeneracy between BHs and ``BH mimickers''. We show that this is not the case for gravastars, because their multipolar structure and their tidal deformability in the BH limit are precisely the same as those of a Kerr BH.

Through this work we use geometrized units where Newton's constant and the speed of light are set to unity; we use $c:= M/R$ to denote the compactness of the object.

\section{Slowly-rotating thin-shell gravastars}
We focus on the simplest gravastar model~\cite{Visser:2003ge}, which consists of an exterior, vacuum solution of Einstein's equation glued to an interior solution supported by a positive cosmological constant (see also Refs.~\cite{Mazur:2001fv,Carter:2005pi,Cattoen:2005he} for different models). The interface defines the object's radius $R$; a material thin shell located at $r=R$ is required because the matching of the two geometries is not smooth~\cite{Mazur:2001fv,Visser:2003ge,Mazur:2004fk,Chirenti:2007mk,Pani:2009ss}.

Following Hartle and Thorne~\cite{Hartle:1967he,Hartle:1968si}, we consider the following ansatz for a slowly-rotating object to second order in the spin,
\begin{eqnarray}
 &ds&^2=-f\left[1+2\left(j_0+j_2 P_2\right)\right]dt^2+\frac{1+\frac{2(m_0+m_2P_2)}{r h}}{h}dr^2\nn\\
 &&+r^2\left[1+2(v_2-j_2)P_2\right]\left[d\vartheta^2+\sin^2\vartheta(d\varphi-\omega dt)^2\right]\,, \label{metric}
\end{eqnarray}
where $P_2:= P_2(\cos\vartheta)=(3\cos^2\vartheta-1)/2$ is a Legendre
polynomial. The radial functions $f$ and $h$ are of zeroth order in
rotation, $\omega$ is of first order, and $j_0$, $j_2$,
$m_0$, $m_2$, $v_2$ are of second order.

Using the ansatz above, it is straightforward to solve Einstein's equations perturbatively for $\chi:= J/M^2\ll1$. To zeroth order in the spin, Birkhoff's theorem guarantees that the exterior solution is identical to the Schwarzschild metric, whereas the interior is described by a patch of de Sitter spacetime. Thus, to ${\cal O}(\chi^0)$, the solution reads~\cite{Mazur:2001fv,Visser:2003ge,Mazur:2004fk}
\begin{equation}
f(r)=h(r)=\left\{\begin{array}{l}
   1-2c {r^2}/{R^2} \qquad r<R\\
   1-{2M}/{r} \qquad \hspace{0.34cm} r>R
  \end{array}\right.\,,
\end{equation}
where $M$ is the gravastar mass measured by an observer at infinity, $c:= M/R$ is the compactness, and the effective cosmological constant of the de Sitter region is $\Lambda:= 8\pi\rho_\Lambda=6M/R^3$. 
The junction conditions at the surface ($r=R$) have already been partially chosen by
requiring the induced metric $g_{ij}$ (where $i,j=t,\vartheta,\varphi$) to be continuous across the shell (cf. Ref.~\cite{Pani:2009ss} for details). 
Israel's junction conditions~\cite{Israel:1966rt} then relate the discontinuities of the extrinsic curvature on the shell with the stress-energy tensor of the thin layer. From these conditions, the surface
energy $\Sigma$ and surface tension $\Theta$ of the shell read \cite{Visser:2003ge}
\begin{equation}
[[\sqrt{h}]] =-4\pi R \Sigma\,,\quad 
\left[\left[ {f'\sqrt{h}}/{f}\right]\right]= 8\pi (\Sigma-2\Theta)\,, \label{eq:SigmaTheta}
\end{equation}
where the symbol $[[A(r)]]:= \lim_{\epsilon\to0} [A(R+\epsilon)-A(R-\epsilon)]$ denotes the discontinuity of a generic function $A(r)$ across
the shell. In this simple model, the coefficient $h$ is continuous across the shell, and therefore $\Sigma=0$, whereas the surface tension is nonzero, $\Theta=-\frac{3c}{8\pi R\sqrt{1-2c}}$. It is easy to see that the thin-shell matter satisfies the weak energy condition but violates the dominant energy condition. Furthermore, the surface tension diverges in the BH limit, as expected by the fact that in this case an event horizon is located at $r=2M$ and the future domain of dependence of the interior has no intersection with the exterior.

Note that the function $f$ can in general be discontinuous if the time coordinate of the external patch is different from the one in the internal patch. This freedom corresponds to different choices of the equation of state of the thin shell. However, for all choices different from the one we made, the corresponding solution has a maximum compactness smaller than that of a Schwarzschild BH (cf. e.g. Ref.~\cite{Uchikata:2015yma}).  Since in this paper we are interested in the approach to the BH limit, we will impose that the time coordinate is the same in the two patches, which singles out the specific equation of state presented above, allowing for models with $c\leq1/2$. A detailed analysis of different equations of state will appear elsewhere~\cite{inpreparation}.

At variance with ordinary stars, gravastars are purely gravitational solutions which are not supported by the pressure of internal fluids. Rotation does not induce any fluid motion and it is thus particularly easy to obtain the first- and second-order spin corrections to the static solution above. To first-order in the spin, the gravitomagnetic term satisfies the following ordinary differential equation
\begin{equation}
 \omega''+\frac{4}{r}\omega'=0\,,
\end{equation}
both in the interior and in the exterior, where a prime denotes derivative with respect to $r$. The general solution to this equation is $\omega(r)=c_0+2c_1/r^3$, where $c_i$ are integration constants. One can fix one of these constants both in the interior and in the exterior by requiring regularity at the center and asymptotic flatness at infinity, respectively. We obtain
\begin{equation}
\omega(r)=\left\{\begin{array}{l}
   \Omega \qquad \hspace{0.15cm} r<R\\
   \frac{2J}{r^3}\qquad r>R
  \end{array}\right.\,,
\end{equation}
where $\Omega:= c_0$ and $J:= c_1$ are now the object's angular velocity and angular momentum, respectively, and are still free constants. However, continuity of the induced metric across the shell implies $\omega$ to be also continuous, and this requirement translates into a relation between $\Omega$ and $J$, namely $\Omega=2J/R^3$. Therefore, to ${\cal O}(\chi)$ the moment of inertia of a gravastar is simply
\begin{equation}
 I:= \frac{J}{\Omega} =\frac{R^3}{2} \,,\label{eqI}
\end{equation}
and reduces to the Kerr value, $I=4 M^3 +{\cal O}(\chi^2)$, when $c=1/2$. 

Let us now compute the ${\cal O}(\chi^2)$ corrections. In this case it is more convenient to discuss the interior and the exterior solutions separately. The exterior geometry is the generic (stationary, axisymmetric) vacuum solution of Einstein's equation to second order in the spin, first derived in Refs.~\cite{Hartle:1967he,Hartle:1968si}, namely
\begin{widetext}
\begin{eqnarray}
 m_0&=&\chi ^2  \left[{\delta m}-\frac{c^4 R^4}{r^3}\right] , \label{ext1}\\
 m_2&=& \frac{\chi ^2}{2 c^2 r R^2} \left[{\delta q} \left(c R
   (cR-r) \left(3 r^2-6 c r R-2 c^2 R^2\right)+3 r^2 (r-2 c R)^2
   {\,\tanh^{-1}}\left[\frac{c R}{r-c R}\right]\right)\right.\nn\\
   &&\left.-\frac{2 c^5 R^5}{r^3} (r-5 c R) (r-2 c R)\right] ,\\
 j_0&=&\chi ^2\frac{c^4 R^4-r^3 {\delta m} }{r^3 (r-2 c R)} ,\\
 j_2&=&\frac{\chi ^2}{2 c^2 r^4 R^2} \left[2 c^5 R^5 (r+c R)+\frac{r^3 {\delta q} \left(c R (cR-r)
   \left(6 c r R+2 c^2 R^2-3 r^2\right)-3 r^2 (r-2 c R)^2 {\,\tanh^{-1}}\left[\frac{c
   R}{r-c R}\right]\right)}{r-2 c R}\right], \label{j2ext}\\
 v_2&=&\frac{\chi ^2}{c R}\left[{\delta q} \left(3 (r-c R) {\,\tanh^{-1}}\left[\frac{c R}{r-c
   R}\right]-\frac{c R \left(3 r^2-6 c r R+c^2
   R^2\right)}{r (r-2 c R)}\right)-\frac{c^5 R^5}{r^4}\right],  \label{v2ext}
\end{eqnarray}
\end{widetext}
where $\delta m$ and $\delta q$ are dimensionless integration constants related to the spin-induced shifts of the mass and of the quadrupole moment of the object, respectively~\cite{Hartle:1967he,Hartle:1968si}.

In the interior, $r<R$, the spacetime is a solution of Einstein's equation with a positive cosmological constant, $\Lambda=6c/R^2$. The solution that is regular at the center reads
\begin{widetext}
\begin{eqnarray}
 m_0&=& 0  \,\\
 m_2&=& -\frac{c^{3/2} R^5 \chi ^2}{96 \sqrt{2} r^2} \left(\sqrt{2c} r R \left(10 c r^2-3 R^2\right)+3
   \left(R^2-2 c r^2\right)^2 {\,\tanh^{-1}}\left[\frac{\sqrt{2c}
   r}{R}\right]\right) a_2  \,,\\
 j_0&=& c^4 R^4 \chi ^2 a_0  \,,\\   
 j_2&=& -\frac{c^{3/2} R^7 \chi ^2}{96 \sqrt{2} r^3 \left(2 c r^2-R^2\right)} \left(\sqrt{2c} r R \left(10 c r^2-3 R^2\right)+3
   \left(R^2-2 c r^2\right)^2 {\,\tanh^{-1}}\left[\frac{\sqrt{2c}
   r}{R}\right]\right) a_2  \,,\\
 v_2&=& \frac{c^{5/2} R^6 \chi ^2}{24 \sqrt{2} \left(2 c r^3-r R^2\right)} \left(\sqrt{2c} r \left(4 c r^2-3
   R^2\right)+\left(3 R^3-6 c r^2 R\right) {\,\tanh^{-1}}\left[\frac{\sqrt{2c} r}{R}\right]\right) a_2  \,,  
\end{eqnarray}
\end{widetext}
where $a_0$ and $a_2$ are integration constants.

Continuity of the induced metric $g_{ij}$ across the shell relates the integration constants of the two solutions above, namely: 
\begin{widetext}
\begin{eqnarray}
 a_0 &=& \frac{c^4 R-{\delta m}}{c^4 (1-2 c) R^5} \,,\\
 a_2 &=& -\frac{96 c^{3/2}}{R^6 \Delta(c)} \left(c \left(14 c^2-3 c-3\right)+3 \left(1-6 c^2+4 c^3\right)
   {\,\tanh^{-1}}\left[\frac{c}{1-c}\right]\right)\,,\\
 \delta q&=& \frac{c^5}{\Delta(c)} \left(2 \sqrt{c} \left(16 c^2-9-6 c\right)-9 \sqrt{2} \left(4 c^2-1\right)
   {\,\tanh^{-1}}\left[\sqrt{2c}\right]\right)\,,  \label{deltaq}
\end{eqnarray}
where
\begin{eqnarray}
 \Delta(c)&=&2 \sqrt{c} \left(c \left(9-12 c+9 c^2+8 c^3\right)-3 \left(3-7 c+6 c^2\right)
   {\,\tanh^{-1}}\left[\frac{c}{1-c}\right]\right)\nn\\
   &&-3 \sqrt{2}
   {\,\tanh^{-1}}\left[\sqrt{2c}\right] \left(c \left(3-6 c-5 c^2+6
   c^3\right)-3 \left(1-3 c+4 c^3\right)
   {\,\tanh^{-1}}\left[\frac{c}{1-c}\right]\right)\,. \label{Delta}
\end{eqnarray}
\end{widetext}
As expected, matching the interior geometry with the exterior geometry yields a solution which is continuous across the shell, 
except for the $g_{rr}$ component whose discontinuity is related to the properties of the shell through the junction conditions. 
The induced metric is continuous but its derivatives are generically discontinuous, the jumps across the shell being related to the physical properties of the latter. Once a model for the thin shell is assumed, its properties and the parameter $\delta m$ are uniquely determined in terms of the discontinuities of the extrinsic curvature~\cite{Israel:1966rt,Visser:2003ge}. A more detailed discussion of the properties of the thin shell for various equations of state is in preparation~\cite{inpreparation}\footnote{Very recently, Ref.~\cite{Uchikata:2015yma} investigated in detail the junction conditions for a slowly-rotating, thin-shell gravastar (see also Refs.~\cite{Kashargin:2011fg,Delsate:2014iia} for some related studies on rotating thin shells in general relativity.)}.

The spin-induced mass quadrupole moment $Q$ of the solution can be extracted from the large-distance behavior of $j_2$, namely $j_2\to Q/r^3$ as $r\to\infty$~\cite{Hartle:1967he,Hartle:1968si}. From Eq.~\eqref{j2ext}, we obtain
\begin{equation}
 Q(c)=\chi^2 M^3 \left(1-\frac{4}{5}\delta q(c)\right)\,, \label{eqQ}
\end{equation}
where $\delta q(c)$ is given in Eqs.~\eqref{deltaq}--\eqref{Delta}.

It is straightforward to evaluate the quadrupole moment~\eqref{eqQ} in the Newtonian ($c\to0$) and the BH ($c\to1/2$) limits, i.e.
\begin{equation}
 \delta q\to \left\{\begin{array}{l}
		     5 -\frac{165 }{14}c+\frac{125 }{49}c^2 \qquad  \hspace{0.35cm} c\to0 \\
                     -\frac{2}{9}\left[\log(1-2c)\right]^{-1} \qquad c\to1/2
                    \end{array}\right. \label{regimesQ}
 \,.
\end{equation}
Note that $Q$ is a monotonic function of $c$: it is negative (i.e. the gravastar is prolate) in the Newtonian limit, vanishes at $c\approx 0.358$, and becomes positive (i.e. the gravastar is oblate) for larger compactness. Finally, $\delta q\to0$ and $Q\to \chi^2 M^3$ in the BH limit, precisely as in the Kerr case. This property seems to be valid also for more generic equations of state of the thin shell, as shown in a recent work~\cite{Uchikata:2015yma}.

\section{Tidal perturbations of a thin-shell gravastar}
The tidal deformability of a self-gravitating object is encoded in the tidal Love numbers, which measure the deformation of the multipole moments of the object induced by an external tidal field (see e.g.~\cite{Murraybook,PoissonWill}). For relativistic, nonspinning compact objects, the tidal Love numbers can be computed through the method developed in Ref.~\cite{Hinderer:2007mb} (see also Refs.~\cite{Binnington:2009bb,Damour:2009vw}). Here we focus the electric tidal Love numbers of a static gravastar; the extensions to the magnetic~\cite{Binnington:2009bb,Landry:2015cva,Delsate:2015wia} and slowly-spinning~\cite{Poisson:2014gka,Pani:2015hfa,Landry:2015zfa} cases are straightforward.

We consider static, polar perturbations of the nonrotating background defined by the metric~\eqref{metric} with $\chi=0$ (i.e. where all functions other than $f=h$ vanish). Following Ref.~\cite{Hinderer:2007mb}, we expand the metric as $g_{ab}=g_{ab}^{(0)}+\delta g_{ab}$, where ($a,b=t,r,\vartheta,\varphi$), $g_{ab}^{(0)}$ is the background geometry and\footnote{Similarly to the background solution, here we assume that the time coordinates in the interior and in the exterior of the perturbed solution are the same. This effectively corresponds to fixing the speed of sound on the shell. Other choices will be discussed elsewhere~\cite{inpreparation}.}
\begin{eqnarray}
 \delta g_{ab}&=&{\rm diag}\left(f H_0^{(\ell)}(r) ,h^{-1} H_2^{(\ell)}(r) ,\right.\nn\\
 &&\left.r^2 K^{(\ell)}(r),r^2\sin^2\vartheta K^{(\ell)}(r) \right)Y_{\ell 0}(\vartheta,\varphi)\,.
\end{eqnarray}
In the above equation $Y_{\ell m}$ are the spherical harmonics and a sum over $\ell$ is implicit. Due to the spherical symmetry of the background, the azimuthal number $m$ is degenerate and can therefore be set to zero without loss of generality. Furthermore, multipoles with different values of $\ell$ decouple. By plugging the perturbation above into the linearized field equations, we obtain $H_2^{(\ell)}=H_0^{(\ell)}$ and\footnote{In order to treat the interior and the exterior solutions at the same time, we have derived the equations for the background quantities $f=h=1-2\hat{M}/r-\hat{\Lambda} r^2/3$, where we introduced the definitions $\hat\Lambda=\Lambda \Theta(R-r)$ and $\hat{M}=M\Theta(r-R)$, $\Theta(x)$ being the Heaviside step function. Therefore, $\hat\Lambda=0$ when $r>R$ and $\hat{M}=0$ when $r<R$. In both cases the equations presented in the main text simplify considerably.}  
\begin{eqnarray}
 K^{(\ell)}(r)&=& \frac{1}{3
   \left(\ell(\ell+1)-2\right) r \left(6 \hat{M}-3 r+r^3 \hat\Lambda \right)}\left[\left(36 \hat{M}^2\right.\right. \nn\\
   &&\left.\left.+18 \left(\ell(\ell+1)-1\right) \hat{M} r-9 \left(\ell(\ell+1)-2\right) r^2\right.\right.\nn\\
   &&\left.\left.+3 r^3 \left(16 \hat{M}+\left(\ell(\ell+1)-4\right) r\right) \hat\Lambda -2
   r^6 \hat\Lambda ^2\right) {H_0^{(\ell)}}    \right.\nn\\
   &&\left.-2 r \left(r^3 \hat\Lambda -3 \hat{M}\right) \left(6 \hat{M}-3 r+r^3 \hat\Lambda \right) {H_0^{(\ell)}}'\right]\,,
\end{eqnarray}
whereas $H_0^{(\ell)}$ satisfies a second-order differential equation
\begin{equation}
 \frac{d^2{H_0^{(\ell)}}}{dr_*^2}-V H_0^{(\ell)}=0\,, \label{eqtidal}
\end{equation}
where
\begin{equation}
 V=\frac{9 r^2}{R^4} \left[\ell(\ell+1) f+\frac{4 \hat{M}^2}{r^2}+\frac{2}{9} r^2 \hat\Lambda  \left(9-r^2 \hat\Lambda \right)\right] ,
\end{equation}
and the tortoise coordinate is defined such that $dr/dr_*:= g(r)=\frac{r}{R^2} \left(6 \hat{M}+\hat\Lambda  r^3-3 r\right)$. Note that the potential $V$ is discontinuous across the radius, whereas $g(r)$ is continuous by virtue of the fact that $\Lambda=6M/R^3$.

Because of such discontinuity, appropriate junction conditions have to be derived for the function $H_0^{(\ell)}$. Since Eq.~\eqref{eqtidal} is homogeneous and linear, we can assume that $H_0^{(\ell)}$ is continuous across the shell without loss of generality. To obtain the discontinuity of ${H_0^{(\ell)}}'$ at the radius, we integrate Eq.~\eqref{eqtidal} from $R-\epsilon$ and $R+\epsilon$ and then take the $\epsilon\to0$ limit. This procedure yields the junction condition
\begin{equation}
 \left[\left[{H_0^{(\ell)}}'\right]\right]= \frac{R}{g(R)}  H_0^{(\ell)}(R) [[V]]\,, \label{junction}
\end{equation}
where we recall that a prime denotes derivative with respect to $r$.  The discontinuity of the potential does not depend on $\ell$, $[[V]]={108 M (M-R)}/{R^4}$, and therefore
\begin{equation}
 \left[\left[{H_0^{(\ell)}}'\right]\right]= 36\frac{c (1-c)}{1-2c}\frac{H_0^{(\ell)}(R)}{R}\,, \label{junction2}
\end{equation}
for any value of $\ell$.

Nonradial perturbations of a static, thin-shell gravastar have been studied in detail in Refs.~\cite{Chirenti:2007mk,Pani:2009ss,Pani:2010em,Cardoso:2014sna}. The perturbation equations in the interior can be solved analytically in terms of hypergeometric functions, even in the time-dependent case~\cite{Pani:2009ss,Pani:2010em,Cardoso:2014sna}. In the static limit, an analytical solution is also available in the exterior, so that the problem of static tidal perturbations of a gravastar can be solved in exact form. Indeed, it is easy to show that the regular solution of 
Eq.~\eqref{eqtidal} reads
\begin{equation}
 H_0^{\ell} = \left\{\begin{array}{l}
                            \alpha_1 \left(\frac{r^\ell \sqrt{\Lambda}}{3-r^2\Lambda}\right) {}_2F_1\left[\frac{\ell-1}{2},\frac{1}{2},\ell+\frac{3}{2},\frac{r^2\Lambda}{3}\right] \quad r<R\\
                            \alpha_2 P_\ell^2(\frac{r}{M}-1)+\beta_2 Q_\ell^2(\frac{r}{M}-1) \quad \hspace{0.875cm} r>R\\
                           \end{array}\right.\,,
\end{equation}
where $P_\ell^2$ and $Q_\ell^2$ are associated Legendre functions, ${}_2 F_1$ is the hypergeometric function, and $\alpha_i$ and $\beta_i$ are integration constants. One integration constant associated to the inner solution has been already fixed by requiring regularity at the center~\cite{Pani:2009ss}. Other two constants (say $\alpha_2$ and $\beta_2$) can be fixed by imposing $\left[\left[H_0^{(\ell)}\right]\right]=0$ and Eq.~\eqref{junction2}, thus leaving a solution which is defined modulo an overall normalization factor, as expected from the linearity and homogeneity of Eq.~\eqref{eqtidal}.

The solution above is valid for generic values of $\ell$. For simplicity, let us focus on the dominant quadrupolar case, fixing $\ell=2$. At large distances, the solution behaves as
\begin{equation}
 H_0^{(2)}\to 3c_1\frac{r^2}{M^2}  -6 c_1 \frac{r}{M}+\frac{8}{5}\frac{M^3}{r^3} c_2+{\cal O}((M/r)^4)\,,
\end{equation}
where $c_1$ and $c_2$ are constants related to $\alpha_2$ and $\beta_2$. After the matching discussed above, these constants are both proportional to the overall normalization factor. Using the same definition adopted in Ref.~\cite{Hinderer:2007mb}, we define the electric, quadrupolar tidal Love number as
\begin{equation}
 \lambda=\frac{8}{5}\frac{c_2}{c_1}M^5\,.
\end{equation}
Note that $\lambda$ depends on the ratio $c_2/c_1$, and therefore the normalization factor cancels out, as expected. Finally, by imposing $\left[\left[H_0^{(\ell)}\right]\right]=0$ and Eq.~\eqref{junction2}, we can explicitly compute the value of $c_2/c_1$, yielding
\begin{widetext}
 \begin{eqnarray}
 \lambda&=& \frac{56 (1-2 c)^2}{\Gamma(c)} \left[10 \sqrt{c} \left(72 c^3-108 c^2+34 c-3\right)+3 \sqrt{2} (1-2 c)^2 \left(36 c^2-40 c+5\right)
   \tanh^{-1}\sqrt{2c}\right]\,,  \label{eqlambda}
\end{eqnarray}
where
\begin{eqnarray}
 \Gamma(c)&=&  210 c\sqrt{c} (720 c^6+440 c^5-5084 c^4+6590 c^3-3210 c^2+645 c-45)+\frac{315 (1-2 c)^2}{2}\nn\\
 &\times&
   \left\{10 \sqrt{c} (6 c-1) (12 c^2-16 c+3) \log(1-2c)+\sqrt{2} \tanh^{-1}\sqrt{2c} \left[2 c (72 c^5+72 c^4-474 c^3+488 c^2-165 c+15)\right.\right.\nn\\
   &&\left.\left.+3 (1-2 c)^2 (36 c^2-40 c+5)  \log(1-2c)\right]\right\}\,.
\end{eqnarray}
\end{widetext}
The result above is exact and valid for any compactness. In the Newtonian and BH limits the tidal Love number reduces to
\begin{equation}
 \frac{\lambda}{M^5}\to \left\{\begin{array}{l}
                    -\frac{18}{7 c^4}\qquad \hspace{1.4cm}  c\to0\\
                    -\frac{128}{9} \left(c-\frac{1}{2}\right)^2 \qquad c\to1/2\\
                   \end{array}\right.\,. \label{regimesL}
\end{equation}
In the Newtonian limit the scaling of the tidal Love number with the compactness is different than in the stellar case, where $\lambda/M^5\sim c^{-5}$~\cite{PoissonWill,Hinderer:2007mb}. Furthermore, at variance with the NS case, the electric tidal Love number is \emph{negative} for any compactness (i.e. an oblate external quadrupolar tidal field makes a spherically-symmetric gravastar prolate and vice versa), and vanishes in the BH limit, i.e. $\lambda\to0$ as $c\to 1/2$. 

These two peculiar properties are captured by a simple Newtonian model. For a self-gravitating barotropic fluid in the Newtonian regime, the quadrupolar electric Love number is governed by the equation~\cite{Hinderer:2007mb}
\begin{equation}
 {H_0^{(2)}}''+\frac{2}{r}{H_0^{(2)}}'+\left[\frac{4\pi\rho}{P'(\rho)}-\frac{6}{r^2}\right]{H_0^{(2)}}=0\,. \label{NewtonianLove}
\end{equation}
Although in dynamical situations the interior de Sitter spacetime is not equivalent to a barotropic fluid, one can attempt at modeling the gravitational perturbations through the equation of state $P(\rho)=-\rho=-\Lambda/(8\pi)$. In such case Eq.~\eqref{NewtonianLove} can be solved analytically and matched to the exterior solution. The same procedure explained above then yields  $\lambda/M^5= -\frac{72}{35} c^{-4}$, which correctly reproduces the scaling, the sign, and also the order of magnitude of the Newtonian Love number in Eq.~\eqref{regimesL}. This suggests that the different scaling of $\lambda$ and its opposite sign relative to the case of an ordinary fluid are due to the peculiar equation of state $P(\rho)=-\rho$ of the de Sitter interior.

We verified that the properties of the electric tidal Love numbers with $\ell>2$ are qualitatively similar to the $\ell=2$ case discussed above, namely $\lambda\sim c^{-2\ell}$ in the Newtonian limit and $\lambda\sim (1-2c)^2$ in the BH limit for any $\ell$.

Finally, the analogy with a Newtonian fluid having equation of state $P(\rho)=-\rho$ is also useful to interpret the results obtained in the previous section for the quadrupole moment, namely that a slowly-spinning gravastar is prolate in the Newtonian limit whereas it becomes oblate approaching the BH limit\footnote{We thank Leo Stein for suggesting this analogy.}. For ordinary bodies, one can think of the oblateness as arising from the object's response to the centrifugal pseudo-force. An object with negative pressure responds in the opposite direction that normal materials responds to forces (i.e., with an anti-restoring force directed as the external force). Since ordinary spinning objects become oblate, one might expect that an object with negative pressure will instead become prolate, in agreement with the Newtonian limit of Eq.~\eqref{eqQ}. Likewise, the fact that a gravastar becomes oblate at high compactness might be interpreted as due to the contribution of the thin shell (made of ordinary matter), which becomes dominant in this limit.

\section{I-Love-Q relations for a gravastar}
Yagi and Yunes~\cite{Yagi:2013bca,Yagi:2013awa} discovered that for NSs and quark stars the dimensionless quantities
\begin{equation}
 \bar I:= \frac{I}{M^3} \,,\qquad \bar{Q}:= \frac{Q}{\chi^2 M^3}\,, \qquad \bar\lambda := \frac{\lambda}{M^5} \,,\label{ILQ}
\end{equation}
satisfy nearly universal relations that are insensitive (at the percent level) to the equation of state of matter in the stellar interior.

\begin{figure}[t]
\begin{center}
\epsfig{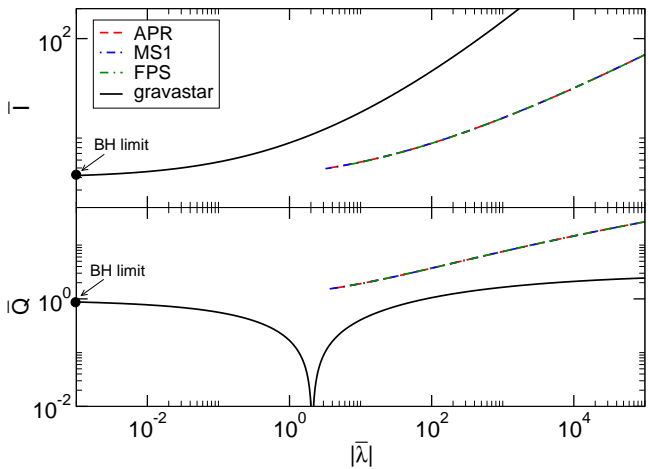}
\caption{The $\bar{I}$-$\bar{\lambda}$ (top panel) and the $\bar{Q}$-$\bar{\lambda}$ (bottom panel) relations for a thin-shell gravastars as computed in this work and compared with those for a NS for various equations of state (see legend) spanning a wide range of NS deformability (cf. Refs.~\cite{Yagi:2013bca,Yagi:2013awa} for details). The BH limit is denoted by a circular black marker. For NSs the curves are truncated at a value roughly corresponding to the maximum mass, whereas gravastar models connect continuously to the BH limit. Note that $\bar{\lambda}<0$ in the gravastar case.
\label{fig:ILQ}}
\end{center}
\end{figure}

For the thin-shell gravastar model under consideration the dimensionless quantities~\eqref{ILQ} can be obtained \emph{analytically} as one-parameter functions depending only on the compactness [cf. Eqs.~\eqref{eqI}, \eqref{eqQ} and \eqref{eqlambda}]. 
These relations are valid in the full range $0\leq c\leq 1/2$, thus connecting smoothly the Newtonian regime to the BH limit.
It is therefore interesting to compare our results with those obtained for NSs and BHs. Figure~\ref{fig:ILQ} shows a comparison of the $\bar{I}$-$\bar{\lambda}$ and the $\bar{Q}$-$\bar{\lambda}$ relations for thin-shell gravastars and for NSs with some realistic equations of state~\cite{Yagi:2013bca,Yagi:2013awa}.

We can observe various interesting features. First, while the relations for NSs are almost insensitive to the equation of state (on the scale of the plot the three curves lay on the top of each other), the same relations for gravastars are dramatically different in the entire range of compactness.  We stress that our gravastar model corresponds to a specific equation of state for the thin shell [cf. below Eq.~\eqref{eq:SigmaTheta}] so that our results do not assess the universality (of absence thereof) of the I-Love-Q relations for a thin-shell gravastar. Nonetheless, such results are sufficient to prove the opposite, i.e., that the I-Love-Q relations of a gravastar are different from those of a NS. This is due to the fact that for gravastars both the low-compactness and the high-compactness regimes are markedly different from the NS case. This is a nontrivial result, since for other compact stellar configurations (e.g. for quark stars) the I-Love-Q relations are the same as those for a NS for a variety of equations of state. 

Secondly, we stress that $\bar\lambda$ is negative for gravastars, i.e. it has the opposite sign than for ordinary stars. As suggested by the Newtonian model discussed above, these peculiar properties are due to the exotic equation of state associated with a positive cosmological constant in the gravastar's interior, $P(\rho)= -\rho$. 

Finally, in the gravastar case the mass quadrupole moment does not have a definite sign, being negative for small compactness, becoming positive for $c\gtrsim 0.358$ (or, equivalently, for $|\bar\lambda|\lesssim 2.09$), and approaching unity in the BH limit. Therefore, slowly-spinning gravastars are prolate for moderately small compactness and become oblate as the compactness increases towards the BH limit. This behavior is qualitatively similar to that observed in strongly-anisotropic, incompressible NSs~\cite{Yagi:2015cda,Yagi:2015hda}.

\subsubsection{The approach to the BH limit}
As already discussed, the BH limit is continuous, namely
\begin{equation}
 \bar I\to4\,,\quad \bar Q\to 1\,,\quad \bar\lambda\to 0\,,
\end{equation}
as $c\to1/2$, precisely as in the Kerr case to quadratic order in the spin. Inverting Eq.~\eqref{eqI}, $c=(2\bar I)^{-1/3}$, and using Eqs.~\eqref{regimesQ} and \eqref{regimesL}, we can easily compute the scaling of the $\bar I$-$\bar \lambda$-$\bar Q$ relations near the BH limit,
\begin{eqnarray}
  \bar Q &\to& 1+\frac{8}{45}\left[\log(1-2(2\bar I)^{-1/3})\right]^{-1}\,, \\
  \bar \lambda &\to& -\frac{128}{9}\left[(2\bar I)^{-1/3}-\frac{1}{2}\right]^2\,,
\end{eqnarray}
as $\bar I\to 4$ (or, equivalently, as $c\to1/2$). In this specific model, the $\bar I$-$\bar \lambda$-$\bar Q$ relations are nonpolynomial functions near the BH limit. This behavior is markedly different from what observed in strongly-anisotropic NSs. In that case, a given multipole moment ${\cal M}$ shows a critical behavior near the BH limit, namely~\cite{Yagi:2015cda,Yagi:2015hda}
\begin{equation}
 \delta:=\frac{{\cal M}}{{\cal M}_{\rm BH}}-1 \sim (1-2c)^k \qquad c\sim 1/2\,,
\end{equation}
where ${\cal M}_{\rm BH}$ is the value of the multipole moment in the BH case and $k$ is a critical exponent which depends only mildly (roughly within $10\%$)
on the equation of state. For anisotropic NSs, the critical exponent for all multipole moments is roughly $k\approx 4$~\cite{Yagi:2015cda,Yagi:2015hda}. On the other hand, for gravastars we obtain $k=1$ for the first current moment, $S_1:= J= I\Omega$, and we even obtain a nonpolynomial behavior for the mass quadrupole moment, $M_2\propto Q$, for which $\delta\sim [\log(1-2c)]^{-1}$.

Finally, we note that for gravastars  $\bar I(c)$, $\bar Q(c)$ and $\bar \lambda(c)$ are monotonic functions of the compactness in the full range $c\in[0,1/2]$, whereas they display a peculiar nonmonotonic behavior in the case of strongly-anisotropic NSs~\cite{Yagi:2015cda,Yagi:2015hda}.

\subsubsection{Eccentricity profiles}
Recently, Ref.~\cite{Yagi:2014qua} proposed a phenomenological explanation for the nearly-universal relations of NSs in terms of an approximate symmetry emerging at high compactness. As the compactness increases, radial profiles of the eccentricity become nearly constant within the star, leading to the emergence of isodensity self-similarity.

\begin{figure}[t]
\begin{center}
\epsfig{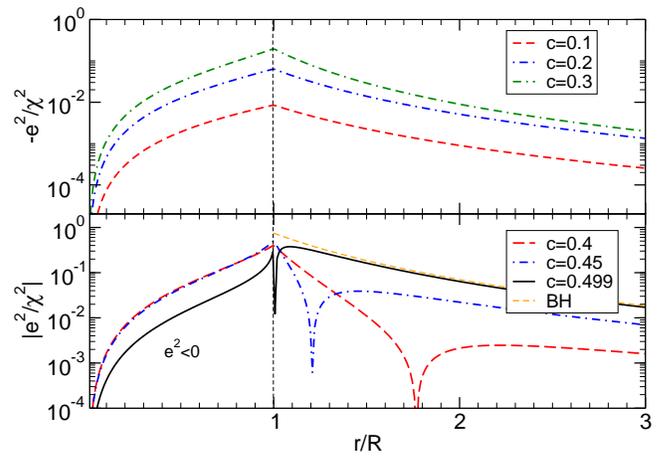}
\caption{Eccentricity profiles of a slowly-rotating, thin-shell gravastar. The top panel shows the profiles for prolate objects ($Q<0$) which requires $c\lesssim 0.358$. In this case $e^2<0$ in the entire space. The bottom panel shows the profiles for oblate objects ($Q>0$). In this case $e^2<0$ up to a certain point in the exterior, and then turns positive for larger values. The BH case, Eq.~\eqref{eccKerr}, is shown as a dashed orange curve. Note that the BH limit is discontinuous because $e^2(R)\to-\chi^2/4$ as $c\to1/2$, whereas $e^2_{\rm BH}(r=2M)=9\chi^2/128$.
\label{fig:eccentricity}}
\end{center}
\end{figure}

It is therefore interesting to check whether the gravastar model enjoys a similar property. For our metric ansatz~\eqref{metric} and in the absence of a fluid, the eccentricity is simply defined as~\cite{Hartle:1967he,Hartle:1968si}
\begin{equation}
 e(r) = \sqrt{-3[v_2(r)-j_2(r)]}\,. \label{eccdef}
\end{equation}
For ordinary objects, $v_2-j_2<0$ when the object is oblate, so that the eccentricity is a real number. In the case of a gravastar instead, $v_2-j_2>0$ in the interior the object for any compactness. 
Using the analytical results previously presented, it is straightforward to obtain an analytical expression for the eccentricity in the entire space. 
The explicit form is not illuminating so we do not show it here, but in Fig.~\ref{fig:eccentricity} we present the eccentricity profiles for various values of the compactness in the entire space.

As anticipated, $e^2<0$ for $r<R$ and any compactness. This is a peculiar feature, because it corresponds to an object which is (locally) prolate, even though the quadrupole moment can be either positive or negative depending on the compactness. Indeed, for $c\lesssim 0.358$ (top panel of Fig.~\ref{fig:eccentricity}) we observe that $e^2<0$ in the entire space, in agreement with the fact that in this case $Q<0$ and therefore the object appears prolate also at infinity. On the other hand, when $c\gtrsim 0.358$ (bottom panel of Fig.~\ref{fig:eccentricity}) we observe that $e^2(r)$ changes sign in the exterior and becomes positive at large distances. In this case, the object has a positive quadrupole moment but a negative eccentricity in the interior. The turning point of the function $e^2(r)$ decreases as the compactness increases and tends to the radius as $c\to1/2$. 

In the BH limit the eccentricity at the radius is discontinuous. Indeed, for our gravastar model $e^2(R)\to-\chi^2/4$ as $c\to1/2$, whereas the eccentricity of a (slowly-spinning) Kerr BH can be easily computed from Eq.~\eqref{eccdef} and by using the external solution~\eqref{j2ext}--\eqref{v2ext} with $\delta m=\delta q=0$. This yields
\begin{equation}
 e^2_{\rm BH}=\frac{3 (r/M+1)}{8 (r/M)^4} \chi^2 \,, \label{eccKerr}
\end{equation}
and the eccentricity at the event horizon is positive, $e^2(2M)=9\chi^2/128$.

In light of these peculiar properties, it is difficult to make a connection to the eccentricity profiles of ordinary stars~\cite{Yagi:2014qua,Yagi:2015hda}, in which $e^2>0$ and the eccentricity remains nearly constant inside the star as the compactness increases. In the gravastar case, the eccentricity vanishes near the center, $e^2\sim r^2$ as $r\to0$, and does not approach a nearly constant profile as $c\to1/2$, although it shows some interesting features that might be worth exploring in more details. In particular, an analysis of the eccentricity profiles for various equations of state of the thin shell is underway~\cite{inpreparation}.

\section{Discussion} \label{sec:conclusions}
The fact that the $\bar I$-$\bar \lambda$-$\bar Q$ relations for gravastars are completely different from those of NSs and quark stars is not surprising. A gravastar is a rather exotic model whose interior is not described by an ordinary fluid. In particular, being a gravastar a purely gravitational object, rotation and tidal fields do not induce any fluid motion. In fact, gravastars have been proposed to mimic BHs rather than compact stars and they share more features with the former, for example in the static case they can have any mass and a compactness $c\to1/2$.

More interestingly, the $\bar I$-$\bar \lambda$-$\bar Q$ relations for gravastars approach smoothly the corresponding value in the (slowly-rotating) Kerr case as the compactness approaches its maximum value\footnote{Note that, because of the slow-rotation expansion, the BH limit corresponds to $c\to1/2$ also when the object is rotating. In this case the effective compactness can be larger due to spin-induced mass shifts. Since we are interested in the dominant spin corrections, the relations~\eqref{ILQ} are normalized by the mass $M$ of the static object and higher-order spin corrections are neglected.}, $c\to 1/2$. 
This fact is remarkable because the BH limit does not need to be continuous and, in fact, the eccentricity at the surface does not connect continuously to its Kerr value as $c\to1/2$. 

The fact that the multipole moments approach their Kerr values has also some phenomenological consequence. It is usually assumed that a measurement of the mass quadrupole moment of a supermassive compact object would be an ultimate test of the BH no-hair theorem and, in turn, of the ``Kerr hypothesis'' (cf. e.g. Ref.~\cite{Berti:2015itd}). Gravastars challenge this paradigm, because they can be as massive and as compact as BHs and, furthermore, they have the same moment of inertia, mass quadrupole moment and even the same (vanishing) tidal Love numbers. 

Although we did not compute higher multipole moments and the magnetic Love numbers, there is no reason to expect qualitatively different results in those cases. It is therefore natural to conjecture that the exterior geometry of a spinning, thin-shell gravastar is \emph{equivalent} to that of a Kerr BH in the $c\to1/2$ limit. This would imply that gravastars and BHs are indistinguishable on the basis of tests probing the multipolar structure of the geometry (e.g. through geodesic motion) or probing the tidal deformability of these objects. Our results suggest that the only direct way\footnote{Broderick and Narayan~\cite{Broderick:2005xa} propose a strong argument which favors the existence of an event horizon (rather than a hard surface) for ultracompact dark objects. In brief, a hard surface would heat up through accretion and even extremely slow accretion rates would be incompatible with the tiny emission from compact dark objects in the galactic centers. This can be regarded as indirect evidence for the existence of supermassive BHs.} to distinguish these objects from BHs is through gravitational-wave observations of the ringdown~\cite{Chirenti:2007mk,Pani:2009ss,Cardoso:2014sna} or of the inspiral in the extreme mass ratio limit~\cite{Pani:2010em,Macedo:2013qea}.

On the other hand --~if one is willing to consider the gravastar proposal seriously~-- the same formation mechanism of BH-like gravastars should also produce stellar-like objects with smaller compactness and with a mass comparable to that of a NS. The existence of different $\bar I$-$\bar \lambda$-$\bar Q$ relations can therefore be used to rule out stellar-like gravastars once independent measurements of any two elements of the triad become available. Figure~\ref{fig:ILQ} shows that, for a fixed value of $|\bar\lambda|$, the deviations of $\bar{I}$ and $\bar{Q}$ for a gravastar relative to a NS are larger than $\sim 200\%$ and $\sim 80\%$, respectively. For moderately compact gravastars, the dimensionless moment of inertia can be larger than that of a NS with same quadrupole moment by more than a factor $10$. Therefore, even a measurement with large uncertainties would easily discriminate between a gravastar and a NS.

Finally, our results provide a simple testbed to explore the approach of the $\bar I$-$\bar \lambda$-$\bar Q$ relations to the BH limit. In this limit the behavior of gravastars is remarkably different from that of anisotropic NSs discussed in Refs.~\cite{Yagi:2015cda,Yagi:2015hda}. In particular, in the BH limit gravastars do not satisfy the critical behavior found for anisotropic stars, either by displaying a different critical exponent or even by displaying a nonpolynomial behavior near the critical point. 
Furthermore, the eccentricity inside the star does not approach a nearly-constant profile as the compactness increases, but displays some peculiar features which are absent in the NS case~\cite{Yagi:2014qua,Yagi:2015hda}.

It would be interesting to extend this analysis to other ``BH mimickers'', e.g. wormholes, quasi-BHs, other black foils (cf. e.g. Refs.~\cite{Lemos:2007yh,Lemos:2008cv}) and to gravastars with different equations of state for the thin shell, to understand whether the approach to the BH limit is characterized by some universal behavior.

\noindent{\bf Note added:} After a first version of this work was submitted, we became aware of a very similar analysis in Ref.~\cite{Uchikata:2015yma}, which also investigates slowly-rotating thin-shell gravastars to quadratic order in the spin. Our work and that of Ref.~\cite{Uchikata:2015yma} are complementary to each other: we investigate the $I$-$\lambda$-$Q$ relations and the approach to the BH limit, whereas Ref.~\cite{Uchikata:2015yma} is devoted to a rigorous analysis of the junction conditions at the interface and on the properties of the thin shell. 
%
\begin{acknowledgments}
I am indebted to Leo Stein for interesting correspondence and for suggesting the analogy with a Newtonian fluid with equation of state $P(\rho)=-\rho$.
This work is supported by the European Community through
the Intra-European Marie Curie Contract No.~AstroGRAphy-2013-623439, by FCT-Portugal through the project IF/00293/2013, by ``NewCompstar'' (COST action MP1304) and by the NRHEP 295189 FP7-PEOPLE-2011-IRSES Grant. 
\end{acknowledgments}
%
%
\bibliography{tidalrot}
\end{document}